\documentclass{article}
\usepackage{slashed}

\def\beq{\begin{eqnarray}}
\def\eeq{\end{eqnarray}}
\def\bea{\begin{eqnarray*}}
\def\eea{\end{eqnarray*}}

\def\centeron#1#2{{\setbox0=\hbox{#1}\setbox1=\hbox{#2}\ifdim
\wd1>\wd0\kern.5\wd1\kern-.5\wd0\fi
\copy0\kern-.5\wd0\kern-.5\wd1\copy1\ifdim\wd0>\wd1
\kern.5\wd0\kern-.5\wd1\fi}}
\def\ltap{\;\centeron{\raise.35ex\hbox{$<$}}{\lower.65ex\hbox{$\sim$}}\;}
\def\gtap{\;\centeron{\raise.35ex\hbox{$>$}}{\lower.65ex\hbox{$\sim$}}\;}

\def\singleandthirdspaced{\baselineskip=\normalbaselineskip\multiply
    \baselineskip by 130\divide\baselineskip by 100}

\newcommand{\newc}{\newcommand}
\newc{\qbar}{{\overline q}}
\newc{\Kahler}{K\"ahler }
\newc{\deltaGS}{\delta_{\rm GS}}
\begin{document}
\begin{titlepage}
\begin{flushright}
{\large hep-th/yymmnnn \\
SCIPP SCIPP 12/12\\
}
\end{flushright}

\vskip 1.2cm

\begin{center}

{\LARGE\bf Gravity Mediation Retrofitted}

\vskip 1.4cm

{\large  Milton Bose and Michael Dine}
\\
\vskip 0.4cm
%{\it $^a$Stanford Linear Accelerator Center,
%     Stanford CA 94309} \\
{\it Santa Cruz Institute for Particle Physics and
\\ Department of Physics,
     Santa Cruz CA 95064  } \\
\vskip 4pt

\vskip 1.5cm

\begin{abstract}
Most discussion of metastable dynamical supersymmetry breaking (MDSB) has focussed on low energy breaking, as in gauge mediation.  It is of interest
to consider possible implications for intermediate scale breaking (``gravity mediation"), especially as
the early LHC results suggest the possibility that supersymmetry, if broken at relatively low energies, might be tuned.  A somewhat
high scale for susy breaking could ameliorate the usual flavor problems of gravity mediation, resolve the question of cosmological
moduli, and account for a Higgs with mass well above $M_Z$.
We study MDSB in gravity mediation, especially retrofitted models in which discrete $R$ symmetries play an important role,
considering questions including implications of symmetries for $B$ and $A$ terms, and the genericity of split supersymmetry.\end{abstract}
\end{center}

\vskip 1.0 cm

\end{titlepage}
\setcounter{footnote}{0} \setcounter{page}{2}
\setcounter{section}{0} \setcounter{subsection}{0}
\setcounter{subsubsection}{0}

%%%%%%%%%%%%%%%%%%%%%%%%%%%%%%%%%%%%%%%%%%%
%%%%%%%%%%%%%%%%%%%%%%%%%%%%
\singleandthirdspaced

\section{Introduction}

If supersymmetry has something to do with the hierarchy problem, it is almost certainly dynamically broken.  First, this is necessary to naturally
account for a large hierarchy.  Second, the ``landscape", whatever its limitations, provides a {\it model} for considering questions of naturalness, and in this
context, if supersymmetry breaking is not dynamical, breaking at the highest scales is favored\cite{douglasdenef,susskind,dinethomas}.

Until the work of Intriligator, Shih and Seiberg (ISS)\cite{iss}, however, dynamical supersymmetry breaking appeared to be a special, almost singular, phenomenon\footnote{In the context of dynamical
models,  metastability had earlier been exploited in in \cite{dimopoulosmetastable}}.  With ISS, the focus
shifted to metastable, dynamical breaking (MDSB), and this appears generic.  The simplest implementation of such breaking occurs in retrofitted models\cite{retrofitted}.
Much of the work on models of MDSB has focussed on gauge mediation.  In the framework of retrofitted models, a number of interesting results have been
 obtained\cite{dinekehayias}:
 \begin{enumerate}
 \item  Models can be constructed with a range of SUSY breaking scales
 \item  One can account for the breaking of the approximate $R$ symmetry of the low energy theory, generating 
 suitable gaugino masses.
 \item  One can naturally account for a small $\mu$ term.
 \item  The size of the superpotential is parametrically of the correct order of magnitude to account for the smallness of the cosmological constant, if the Planck
 scale controls the size of higher dimension operators.
 \end{enumerate}

 As the LHC has already excluded significant parts of the supersymmetry parameter space,
 conventional ideas of naturalness are arguably under some stress.  While it is possible that we will find evidence for a natural structure, such as supersymmetry
 with light stops and an additional singlet field, it is
also possible that naive notions of naturalness are simply
not correct.  At an extreme level, the cosmological constant problem, coupled
with a landscape hypothesis,
suggests that perhaps one should abandon notions of naturalness entirely.  Even within a landscape framework,
however, features of parameter distributions and possible competing anthropic pressures might yield a more moderate
degree of tuning, perhaps accounting for scales of supersymmetry breaking of order $10$'s to $1000$'s of TeV.  This 
would be consistent with the mass of the (candidate) Higgs and current supersymmetry exclusions.
High scales might ameliorate or eliminate the problem of flavor changing neutral currents at low energies and
the cosmological moduli problem\footnote{The virtues of high scales, and their associated
tuning, for the moduli problem were noted in \cite{bkn, deCarlos:1993jw},
and more recently have been stressed in the connections noted here by\cite{kane},\cite{douglas} and \cite{nima}}.
It is then interesting to reconsider models of ``gravity mediation".  It is hard to see how such models could account for hierarchies unless
the breaking is dynamical. The possibility that supersymmetry is dynamically broken in supergravity models in a stable 
vacuum has been considered for some time\cite{ads,dinemacintire,dinesupergravitydsb}, and more recently in \cite{yanagida1,yanagida2,yanagida3}.
We will review features of such models, and their virtues and limitations, in section \ref{stabledsb}.  

Our focus in this paper, however, will be on metastable dynamical supersymmetry breaking in the context of
supergravity,
and especially on retrofitted models.
In retrofitted models, symmetries and their
dynamical breaking play a central role, and it is possible that considering such theories will lead to new insights
into longstanding puzzles.   Questions such as the notion of an approximate, continuous $R$ symmetry, the generation of gaugino masses,
and the $\mu$ term may be seen in a different light than in the past. 
It is also possible that these sorts of considerations might point in new directions for a more natural phenomenology.   Exploring these
possibilities is the goal of the present paper.

Discrete R symmetries, spontaneously broken, are a feature of retrofitted models.  Such symmetries can account for
the approximate, continuous $R$ symmetries required by the Nelson-Seiberg theorem; they can also help account for the smallness
of the cosmological constant.  So we will assume such symmetries throughout this paper. We will ask whether these lead to restrictions
on the soft breaking parameters at low energies.   These symmetries can forbid, not only a large $\mu$ term, but also
the Giudice-Masiero coupling\cite{giudice}; at the same time, other sources of $\mu$ can arise naturally
within this structure.  For the $A$ and $B$ terms we will see that there are significant constraints in certain circumstances and not in
others.  Perhaps most interesting, these symmetries control whether gaugino masses are generated at tree level or in loops.
This is particularly relevant to assessing the genericity of ``split supersymmetry"\cite{split}.  It is often argued that
split supersymmetry is natural, as symmetries can protect gaugino masses, but not scalar masses.  A symmetry under 
which the gauginos transform is necessarily an $R$ symmetry, however, and the scale of $R$ symmetry
breaking is tied to the cosmological constant.  In retrofitted models, this correlation, at the order of magnitude level,
can be natural.   In these cases, the Goldstino supermultiplet (the chiral multiplet whose fermonic
component is the longitudinal mode of the gravitino\footnote{As we will discuss, this notion is not always sharp; we will
clarify when needed.}) is neutral under the $R$ symmetry, and (except in special
circumstances which we will describe) is allowed by all symmetries to couple directly to the gauge fields.  In such cases,
the gaugino masses are typically of order the gravitino mass (as are those of the scalars).  These couplings
may vanish by accident.   Alternatively, as we will describe,
there are theories in which the Goldstino supermultiplet
transforms under a non-R discrete symmetry (and is neutral under the $R$ symmetry).  These theories require additional
features (fields and interactions) to account for moduli stabilization.  Finally, the Goldstino may transform under the $R$ symmetry so that the gaugino masses are
suppressed, at the price, again, of additional interactions and now also {\it very} small parameters.
So we will see that  ``split supersymmetry", while plausible, does not appear particularly
generic.
Even if not, such a phenomenon still may arise by accident, or as a consequence of anthropic tunings.

In section \ref{stabledsb}, we review features of supergravity models which exhibit stable dynamical supersymmetry breaking, and contrast
with MDSB, both for the ISS models and retrofitted theories.
In section \ref{role}, we turn to MDSB, explaining in more detail why one might expect a role for discrete $R$ symmetries.  We discuss why the breaking
should be small, and why there is, as a result, a low energy effective field theory, which to a first approximation
is globally supersymmetric and R symmetric.  In section \ref{neutral}, we discuss models
in which the ``goldstino multiplet" is neutral under the $R$ symmetry.  We will consider mechanisms for
stabilization of the moduli.  We will see that the $B_{\mu}$ and $A$
terms are not predicted in such models (though in some cases $A$ parameters are proportional to Yukawa couplings).  In these models, gaugino and scalar masses are typically of the same order.
In section \ref{charged},
we consider the case that the goldstino multiplet is charged (as we describe, any $R$-neutral moduli might
be fixed at some high scale by additional dynamics).  We will explain the need for additional interactions and small parameters.    In such
a model, the gaugino masses are automatically suppressed by a loop factor.  Predictions for $A$ and $B$ terms can emerge in such a framework.  
Possible origins for fine tuning are discussed in the concluding section (\ref{conclusions}).

\section{Stable vs. Metastable Dynamical Breaking and Supergravity}
\label{stabledsb}

Models of stable, dynamical supersymmetry breaking have been known for some time\cite{ads}.  They have certain characteristic features:
\begin{enumerate}
\item \label{chargedf}  At the level of the lowest dimension operators, they exhibit continuous global symmetries, which are spontaneously
broken.  Typically, fields with non-zero F components carry charges under the corresponding symmetries.
\item\label{pseudomoduli}  There are no approximate flat directions (pseudo moduli).
\item  In renormalizable theories, there is one characteristic scale.
\end{enumerate}  
When coupled to supergravity, these theories have the features that:
\begin{enumerate}
\item  Because of point \ref{chargedf} above, gauginos cannot gain mass from dimension five couplings to fields with
non-vanishing $F$ components.  The leading masses are ``anomaly mediated"\cite{dinemacintire}.
\item  Because of point \ref{pseudomoduli}, there is no moduli problem in these theories.
\item  One requires a large constant in the superpotential, $W_0$.  One can imagine that this is added by hand.
A more principled position is that that it 
arises in a landscape, where there is some continuous distribution of such constants, and anthropically
selected.  Alternative (again in a landscape) it might be generated by
some additional dynamics (and again anthropically selected).
\item  One needs additional features to understand $\mu$.  The Guidice-Masiero mechanism is not operative in these theories.
$\mu$ might be added by hand (again, perhaps, anthropically selected) or be generated by some additional dynamics\footnote{In
\cite{yanagida1}, $\mu$ is generated by a term in the Kahler potential, $c~H_uH_d$.  In the presence of a non-zero $W_0$, and
for $c \sim 1$, this is equivalent to a bare $\mu$ term, as can be seen by performing a Kahler transformation.  If the smallness of
$\mu$ is accounted for by a spontaneously broken $R$ symmetry, say, due to gaugino condensation in another group (also accounting
for $W_0$), this is equivalent to a $W_\alpha^2 H_u H_d$ coupling.} 
\end{enumerate}
Reference \cite{yanagida3} revisits these questions, {\it assuming} that in fact there is a tuned hierarchy of scales, and studies
the phenomenology of these models.

The retrofitted models discussed in the following sections provide a different viewpoint
on many of these issues.  The retrofitted theories typically {\it do} involve supersymmetry breaking
by pseudo moduli.  Symmetries (generally discrete $R$) are inherent in these models.  They have several
promising features:
\begin{enumerate}
\item  In a broad class of models,
the dynamics automatically generate a constant in the superpotential of the required order of magnitude (as we will review
in the next section) to yield a small cosmological constant.
\item  They contain symmetries which suppress the $\mu$ term.
\item  The same dynamics which generates supersymmetry breaking and the constant in the superpotential can
generate a $\mu$ term; alternatively, the Guidice-Masiero mechanism may be operative.
\item  The models suffer from a moduli problem, but this may be a positive feature:  the required large mass for the modulus
may have an anthropic origin (accounting for tuning -- and a lower bound on the susy-breaking scale).
\end{enumerate}

In the context of gravity mediation, the retrofitted models are distinguished from the ISS models.  Indeed, the original ISS models are
closer to the models with stable supersymmetry breaking in structure.  They don't possess moduli; they require an additional
constant in the superpotential, or some new dynamics, to account for the smallness of the cosmological constant; absent the
constant in the superpotential, they
typically have approximate $R$ symmetries which prevent a gaugino mass, so the anomaly mediated contributions
dominate.  Typically additional dynamics is necessary to account for the $\mu$ term.

\section{The Role of Discrete $R$ Symmetries}
\label{role}

There are several reasons why we might expect discrete R symmetries to play a role in any understanding of supersymmetry breaking.
The first has to do with the cosmological constant.
In order to understand the smallness of the cosmological constant, it is necessary that
\beq
W_0 = \langle W \rangle
\eeq
be small.  The only {\it natural} way to understand this is to suppose that there is an underlying, {\it discrete}, R symmetry.  Of course, we do not have a natural understanding
of the dark energy overall, and one might simply view the smallness of $W$ as arising as a part of some anthropic selection of small cosmological constant.
This is the assumption of most landscape analyses\cite{douglasdenef}.  But in a landscape, if both $W_0$ and the scale
of supersymmetry breaking are dynamically generated, the overall level of fine tuning might be significantly
reduced\cite{dgt}.  In retrofitted models, the small breaking of the $R$ symmetry induces SUSY breaking of the order required to give
small c.c.
A second reason to consider discrete $R$ symmetries is the requirement of an (approximate) continuous $R$ symmetry to account for the spontaneous
breaking of supersymmetry (in a metastable vacuum).  Such a symmetry might arise as a result of accidents involving the structure of the gauge-invariant renormalizable
couplings in a theory, but it could also arise from the restrictions on the structure of low dimension operators imposed by an $R$ symmetry; for example, the
discrete symmetry might be a subgroup of the approximate continuous symmetry.

Such symmetries might be relevant, as well, to understanding proton stability and other issues in supersymmetric theories.  Issues
with understanding such symmetries in a landscape context have been discussed in \cite{dinesun,dinefestucciamorisse}, with counting of states
in explicit models performed in several explicit constructions\cite{DeWolfe:2004ns,ibanezetal}.  There it was noted that in flux landscapes,
discrete symmetries are rare, but a picture in which cosmology might favor such symmetries was put forward.
We will assume the presence of such symmetries in what follows.

Given the assumption that there is an underlying discrete $R$ symmetry, 
the first question we might ask is:  should we impose anomaly constraints?  Model builders often
demand satisfaction of some putative set of discrete anomaly constraints.  It is well known, from studies of string theory\cite{banksdinediscrete} that, until one commits oneself to the structure of the microscopic
theory (e.g. a conventional grand unified theory) one can demand,
at most, the cancellation of
anomalies associated with non-abelian gauge groups.  But even for these, if there
are light scalars, anomalies can be canceled by a Green-Schwarz mechanism.  In heterotic string examples, when this occurs, one
often finds that
all anomalies can be cancelled by couplings to a single field\cite{dinegraesser}.%\footnote{One
%{\it can} construct an argument in the case that these other moduli exhibit points of enhanced symmetry.  This will be discussed elsewhere}
A priori, in the presence of multiple moduli, anomalies not only need not vanish, but need not be equal\cite{Blumenhagen:2005pm,DineMonteux2012}.
But there is a simpler, more macroscopic reason, that one should not impose anomaly constraints. 
Any such $R$ symmetry is necessarily broken at a high
scale, given the small value of the observed cosmological constant.  It is possible that fields transforming
under the discrete symmetry gain mass at this scale.  If the breaking of the symmetry is dynamical,
and if, in first approximation, supersymmetry is unbroken, possible order parameters for this breaking
include, as discussed in \cite{dinekehayias}, gaugino condensates, of dimension three, and scalar fields, of dimension one, associated
with some new gauge group.  So such masses can be far larger than $m_{3/2}$.  
As a result of these considerations, we do not view anomalies as constraining\cite{DineMonteux2012}.%\footnote{One might wonder whether unification
%imposes constraints; however, it is possible to write models, consistent with unification, in which the massive states
%contribute to an anomaly.  This will be discussed elsewhere.}  

\section{Retrofitting the Polonyi Model:  Neutral Fields}
\label{neutral}

In this section we will make the simplifying assumption that the only scale in the microscopic theory is $M_p$.
We will also assume that the theory consists of a gauge theory which breaks a $Z_N$ $R$ symmetry without breaking supersymmetry; an
$SU(N)$ gauge theory without matter fields provides a
simple example, but others have been explored in \cite{dinekehayias, kehayias}.

A simple model for supersymmetry breaking in supergravity then consists of a single field, $X$, neutral
under the $R$ symmetry, and  coupled to a supersymmetric gauge theory, with coupling
\beq
W = -{1 \over 4}  {f({X \over M_p}) } W_\alpha^2,
\eeq
where $W_\alpha$ are the gauge fields of the $R$-breaking sector. 
By a holomorphic redefinition of the fields, we can take
\beq
W = -{1 \over 4} \left ( {1 \over g^2} + a {X \over M_p} \right )  W_\alpha^2,
\label{xcoupling}.
\eeq
Because $X$ is neutral under {\it any} symmetry of the theory, the definition of the origin is arbitrary.  Moreover, $X$ is a pseudo modulus,
in that no couplings of the form $X^n$ are permitted by the symmetries.  String theory
models would suggest that $X$ might transform under an approximate shift (Peccei-Quinn) symmetry, $X \rightarrow X + i \alpha$.
Non-perturbative effects would generate a small,
non-perturbative (explicit) breaking of the symmetry.

The interaction of eqn. \ref{xcoupling} leads to a superpotential for $X$:
\beq
W(X) = \Lambda^3 e^{-{ {X \over b M_p} }}
\eeq
for some constant $b$.  $\Lambda$ is the scale of the hidden sector
dynamics, at the (arbitrarily chosen) $0$ of $X$.  An alternative is to define $X$
so that 
\beq
\langle X \rangle = {- {1 \over g^2}} + \dots
\eeq
($g^2 = g^2(M_p)$).
This allows us to write 
\beq
W(X) = M_p^3 e^{-{ X/b }}
\eeq
$W(X)$ yields a potential for $X$, which vanishes for large $X$ as $e^{-2 {\rm Re} ~X/b}$.  By assumption, the potential
has a (metastable) minimum.
$X$ may be stabilized by features of the Kahler potential
(``Kahler stabilization"), described in section \ref{stabilization}, or as a result of couplings to fields which became massless at points on the moduli space. 
The latter is necessary in models of low energy (gauge-mediated) supersymmetry breaking\cite{dinekehayias}.

As noted in \cite{dinekehayias}, with these choices of scalings, vanishing of the cosmological constant can arise
if $a$ is an ${\cal O}(1)$
number (albeit adjusted to many decimal places).  We will see that this is not the case for other possible
mechanisms for supersymmetry breaking.

The underlying theory may contain multiple fields like $X$, neutral under the $R$ symmetry.  It might contain charged moduli as well.
Ignoring the latter, for the moment, we can label the neutral fields by $X_i$, $i=0,\dots N$, and define $X_0$ so that 
\beq
\langle F_i \rangle = 0~~~i > 0.
\eeq
From the perspective of symmetries, $X_0$ is not distinguished in any particular way.

\subsection{Soft Breakings:  Moduli Stabilization}
\label{stabilization}

When considering soft breakings, the first question one needs to address is the stabilization of the modulus (moduli) $X$.
Neutral moduli might be stabilized by features of the Kahler potential, ``Kahler stabilization"\cite{kahlerstabilization}.  It will
be useful to be explicit about what this means.  For a single field, we can simply define $X=0$ as the stationary point of
the potential, as in eqn. \ref{xcoupling}.  Then we can write a Taylor series expansion of $K$:
\beq
K = k_0 + k_1 X  + {\rm c.c.} + k_2 X^\dagger X + \tilde k_2 X^2 + {\rm c.c.} + k_3 X X^{\dagger ~2} + \tilde k_3 X^3 + {\rm c.c.}
\eeq
We impose the conditions
\beq
V^\prime(0) = V(0) = 0.
\eeq
These are two algebraic conditions on the $k_i$'s; they have a multi-parameter set of solutions.  There is no small parameter in these
equations, and the $k_i$'s (in Planck units) generically are comparable.

For the question of gaugino masses, we will be interested in
\beq
\langle F_X \rangle = {\partial W \over \partial X} + {\partial K \over \partial X} W =\Lambda^3 (-{1 \over b} + k_1) .
\eeq

.

\subsection{Soft Breakings:  Gaugino Masses}
\label{gauginomass}

It is often remarked that gaugino masses may be small compared to squark and slepton masses, as a result of the
chiral symmetries which can protect fermion masses.  Any symmetry under which gauginos transform would necessarily be an $R$ symmetry,
and this symmetry, in turn, is necessarily broken, given the smallness of the cosmological constant, which
requires a non-zero expectation value of the superpotential, $W_0 = \langle W \rangle$.   The scale
of this breaking is tied to the scale of supersymmetry breaking, $W_0 = m_{3/2} M_p^2$.  $W_0$,
at the very least, gives rise to the anomaly mediated contribution to the masses\cite{anomalymediation1,anomalymediation2}.
The corresponding loop suppression, in such a case, gives rise to the idea of ``split supersymmetry"\cite{split}.\footnote{The
authors of \cite{split} contemplated very large hierarchies between scalar and gaugino masses; they have dubbed this
one-loop hierarchy ``mini-split".}

But given that the $R$ symmetry must be broken by some dynamics, there are potentially other contributions, which may
not be loop suppressed.
Gaugino masses can arise from a $X W_\alpha^2$ coupling, where $W_\alpha$ now refers to the standard model fields.  
Such couplings to the hidden sector are typical of retrofitted models (eqn. \ref{xcoupling}); it would be surprising if similar couplings
to the standard model gauge fields were absent.  We will glean some insight into this question when we consider unification,
below.

The coupling $X W_\alpha^2$ leads to a gaugino mass
\beq
m_{1/2} = F_X k_2^{-1}.
\eeq
We have seen that, once $X$ is stabilized, its $F$-component is of order $m_{3/2} M_p$, and $k_2 = {\cal O}(1)$.
 n general. If $X$ is neutral, this coupling can not be forbidden by symmetries, the gaugino masses at the high scale are of order
 $m_{3/2}$.
As we have remarked, it is possible that couplings of $X$ to the standard model gauge groups
vanish; in this case, the ``anomaly-mediated" contribution dominates for the standard model gauginos.
Still, split supersymmetry, in the framework of a goldstino multiplet
neutral under the $R$ symmetry, would not appear generic.  It might, of course, simply arise
by accident, or it might be selected by requirements for suitable dark matter
or other (anthropic?) constraints.  We will see in section \ref{charged} that under special circumstances, the Goldstino multiplet may be 
charged under non-R symmetries, allowing a natural suppression of gaugino masses.

\subsubsection{Unification}

If there is one such field, defining $X$ as above, unification requires that $X$ couple in the same way to each of the Standard Model groups; one expects that it couples to the additional strong group, with a coupling which might differ by an order one factor.
So in the retrofitted framework with neutral fields, one expects gaugino masses of order scalar
masses.    If there are multiple neutral moduli, as is familiar in string theory, then unification would seem to be a significant constraint.  
%these masses would be of the same order as the scalar and moduli masses.  With multiple moduli, things could be more complicated.
One possibility, simple to describe but not necessarily to realize, is the following:  two
moduli, $X_0$ and $X_1$, where $X_0$ couples only to the hidden sector gauge group, while $X_1$ couples only to standard model fields.  $X_0$ is the Goldstino multiplet; $X_1$ is another neutral multiplet.
This would be consistent with unification, and with an ``anomaly mediated" origin for the gaugino masses for the MSSM.  Any of these scenarios
has implications for the moduli problem of supersymmetric cosmology, as we will discuss in the conclusions.

It is worth noting that in the heterotic string theory compactified on an $R$ symmetric
space, familiar, $R$ neutral moduli are the model-independent dilaton and the radion.  The radion typically couples
in loops to the gauge fields, in a non-universal fashion.

\subsection{Soft Breakings:  $\mu$, $B_\mu$ and $A$ Terms}

As in general supergravity models, there are a variety of sources for masses for the scalar partners of the quarks and leptons, as well as the Higgs
scalars.   We will denote these fields generically by
$\phi_i$.  Writing the terms in the Kahler potential in the form
\beq
K(X,X^\dagger,\phi^i, \phi^{i\dagger})= f(X,X^\dagger) + g(X,X^\dagger)_{ij} \phi^i \phi^{j~\dagger} + \dots
\eeq
allows a completely general
matrix for the $\phi_i^* \phi_j$ soft breakings.  

For the $\mu$, $B_\mu$, and $A$ terms, one might ask whether the $R$ symmetries yield interesting restrictions.
If the product $H_u H_d$ is neutral under the symmetry, then a $\mu$ term is forbidden above the scale of $R$ breaking.
A $\mu$ term {\it can} arise from the familiar Giudice-Masiero mechanism\cite{giudice}:
\beq
{\cal L}_{\mu} = {1 \over M} \int d^4 \theta f(X,X^\dagger) H_u H_d
\eeq
of order $m_{3/2}$.  By rescaling of the Higgs fields, we can take the coefficient of $H_u^* H_u$ and
$H_D^* H_D$ in the Kahler potential to be unity.  In this case, certain
universal contributions to $B_\mu$ terms arise from the terms in the supergravity action:
\beq
V_{sugra} = e^{K}\left ( {\partial W_{eff} \over \partial H_u} {\partial K \over \partial H_u^*} W_0^* - 3 \vert W \vert^2 \right )
\eeq 
However, the term
\beq
{\partial W^* \over \partial X^*}g^{X,X^*}{\partial K \over \partial X} W + {\rm c.c.}
\eeq
depends on ${\partial K \over \partial X}$, which is not constrained by the symmetries.  So there is no prediction for the 
relation between $B_\mu$ and $\mu$.

Similar issues arise for the $A$ terms.  Non-calculable contributions arise from
\beq
{\partial W^* \over \partial X^*}{\partial K \over \partial X} W~g^{X X^*}.
\eeq
While these are proportional to the Yukawa couplings in the superpotential, non-proportional terms would
arise from terms in the Kahler potential of the form:
\beq
\delta K = \gamma_{ij} X^\dagger \phi_i^* \phi_j + {\rm c.c.}
\eeq
These and similar terms might be forbidden by symmetries, yielding $A$ terms proportional to Yukawa couplings.

If $H_u H_d$ carries a non-trivial $R$ charge, not equal to that of the superpotential, the $\mu$ term must be
generated in a different fashion.  If there are singlet fields, $S$, of suitable charge, with $S \ne 0$, couplings
\beq
\kappa {S^{n} \over M^{n-1}} H_u H_d
\eeq
Give rise to a $\mu$ term\cite{dinekehayias}.  Again, however, $B_\mu$ is not predicted without further knowledge
of the microscopic theory; there is a contribution
proportional to ${\partial K \over \partial X}W$.  Similar remarks apply to the $A$ parameter
in this case.

\section{Generalizations of the Polonyi Model:  Charged fields}
\label{charged}

If the Goldstino field were charged under a symmetry, one could suppress the gaugino mass.  Given that, at least
in known string models, there are usually moduli of $R$ charge zero, it is first necessary to ask how theses might
be stabilized.  Given our basic assumption of an underlying $R$ symmetry, the ``KKLT" mechanism\cite{kklt} is not 
available to us, but Kahler stabilization again can provide a solution.
For example, if
\beq
W \approx e^{-N/b}
\eeq
then supersymmetry is unbroken for $N_0$ such that
\beq  {\partial K \over \partial N_0} = -1/b.
\eeq
For self consistency, this must occur for large $N$.  

Suppose that the Goldstino field, $X$, carries a charge under the discrete symmetries.
Then
 $X$ does not couple directly to the gauge fields.  The leading contribution to
gaugino masses arises from the so-called anomaly mediated affects at one loop.  
This is, indeed,  an implementation of the slogan of \cite{split}, that gaugino masses can be suppressed relative
to scalar masses as a result of symmetries.  Interestingly, it is not the $R$ symmetry or any symmetry
carried by gauginos which is responsible -- the suppression of the coupling arises precisely because
the gauginos are neutral under the symmetry.

In order to achieve a model of this type, we must suppose that the $R$ symmetry is broken by a model such as that of \cite{dinekehayias}, where there are order parameters
of dimension one, $\Phi$, with less trivial discrete charges.  For the models of \cite{dinekehayias}, $\Phi^3$ carries
charge $2$ under the
$R$ symmetry. $\Phi = {\cal O}( \Lambda)$, the scale of the underlying gauge dynamics, and $W_0 \sim \Lambda^3$. Then, for example, there may be a superpotential:
\beq
W =  \kappa X {\Phi^{3+n} \over M_p^{n+1}}
\label{wone}
\eeq
Consider, first, $n \ne 0$.
$X$ now carries a non-trivial $R$ charge.  There is a well-defined notion of origin, and there is a meaningful
sense in which $X$ may be small. We will assume for the moment that $X$ is stabilized near the origin;
we will consider the problem of stabilization
shortly.  If this is the entire content of the theory, the cosmological constant is problematic.
The scale of supersymmetry breaking is
\beq
F_X = \kappa {\Lambda^3 \over M_p} \left ( {\Lambda \over M_p} \right )^n.
\eeq
But $\langle W \rangle$ is also of order $\Lambda^3$,
so
\beq
\kappa \sim \left ( {M_p \over \Lambda } \right )^n.
\eeq
This makes sense for $n < 0$, but requires that $\kappa$ is {\it extremely} small.  For $n>0$, additional dynamics are
required to break the $R$ symmetry in a way that can yield a small cosmological constant.

The case $n=0$ is similar to the $X W_\alpha^2$ case of the previous section. $X$ is neutral under $R$ symmetries.  One
now has a natural understanding of the order of magnitude of $W_0$ (i.e. $\kappa \sim 1$).  But one would like to explain the
absence of the $X W_\alpha^2$ coupling.  This requires that $S$ couple to a combination of fields carrying some charge,
preferably a discrete (non-R) charge.  The models of
\cite{dinekehayias} have the
feature that they may exhibit such symmetries.  In particular, these models have multiple singlet fields, $S_i$.
These appear coupled to ``quark" and ``antiquark", fields, $Q_f$ and $\bar Q_f$, transforming as $N$ and $\bar N$
of $SU(N)$.  A simple model with an additional symmetry is
\beq
W = y_1 S_{-2} \bar Q_1 Q_1 + S_1 ( y_2 \bar Q_0 Q_{-1}  + y_3  \bar Q_{-1} Q_0) + \lambda S_{-2} S_1 S_1.
\eeq
In the limit that $\lambda$ is smaller than the other couplings, $y_i$, one can integrate out the $Q$'s, obtaining
an effective superpotential:
\beq
W_{eff} = (y_1 y_2 y_3 S_{-2} S_1 S_1)^{1/N}\Lambda^{3-3/N} + \lambda S_{-2} S_1 S_1.
\eeq
This model has, at the level of dimension four couplings, a $U(1)$ symmetry under which $S_i$'s have charges corresponding to their
subscripts.  The problem is rather general; it is difficult, with only dimension four couplings, to obtain discrete symmetries apart from $Z_3$.
In the event that there are approximate $U(1)$ symmetries, one obtains a one  (complex) dimensional set of vacua and corresponding 
pseudo moduli.  Theses directions may be lifted by higher dimension operators or supersymmetry-breaking effects but they certainly
yield new complications for model building.  So, while it is possible to construct models of this type, they don't appear especially
generic.

%Models with higher dimension couplings suffer from a different cosmological constant problem; now $W$ is parametrically
%so large that the
%ground state is AdS.  Additional dynamics must arise in the theory allowing cancellation of this large contribution to $W$.

So far we have not discussed the stabilization of the modulus $X$.  We require, not only stability and small c.c., but also that
$X \ll M_p$, in order that the symmetry be effective.  This last aspect is problematic, requiring that the models possess
additional features.  The difficulty is that the potential for $X$ is inherently symmetry breaking; there is no
small scale unless it arises from some {\it other} dynamics.  If we suppose that $X$ is stabilized by features of the Kahler potential,
then, unless there are large dimensionless ratios, $X \sim M_p$.  This can be avoided if $X$ couples to other light fields, providing,
essentially, a retrofitted O'Raifeartaigh (as opposed to Polonyi) model.  This requires new fields and additional mass terms.
It is possible to build models along these lines and we will assume such a structure in what follows.

With a superpotential of the form of eqn. \ref{wone}, and with $F_x \sim {W \over M_p}$, scalar masses are of order $m_{3/2}$.
$\phi_i^* \phi_j$ type terms of a completely general form arise due to the terms in the Kahler potential
\beq
\Gamma_{i j} X^* X \phi^{i*} \phi^j.
\eeq
Parametrically, these masses are of order $m_{3/2}^2$.

The Guidice-Masiero Kahler potential gives rise to a $\mu$ term, again, if the charges of $H_u$ and $H_d$ are suitable.  Now,
because $\langle X \rangle$ is small (as is ${\partial K \over \partial X}$), there are only a few sources of $B_\mu$ and $A$ terms in the supergravity
lagrangian.   $B_\mu$ is
then determined in terms of $\mu$ and $m_{3/2}$:
\beq
B_\mu = -m_{3/2} \mu.
\eeq
Similarly, because of the symmetries, the $A$ terms vanish at tree level and leading order in $\Lambda/M_p$.

Alternatively, the Guidice-Masiero term in the Kahler potential might be forbidden by symmetries, and the $\mu$ term
arise as a result of retrofitting or some other mechanism\cite{dinekehayias}.  In that case, one again has a prediction for
$B_\mu$ and $A$.  Again,
\beq
B_\mu = - m_{3/2} \mu~~~A=0.
\eeq

To summarize, the possibility of charged moduli is interesting from the perspective of relatively light
gauginos.  It comes at a price, however. 
\begin{enumerate}
\item  Additional dynamics are required to fix any neutral
moduli without breaking supersymmetry.  (This is similar to the DSB and ISS theories).
\item {\it Extremely} small couplings
are required to fix the charged moduli, while at the same time obtaining small cosmological
constant, if $n < 0$.   
\item New dynamics (possibly related to those which fix neutral moduli) are required to obtain
a small cosmological constant if $n >0$.  (This is similar to the stable DSB and ISS theories).
\item  The case $n=0$ requires additional symmetries beyond the $R$ symmetries, and, except for $Z_3$'s, these introduce additional
pseudo moduli and associated complications.
\item  In all of these cases, additional degrees of freedom (similar to those of O'Raifeartaigh models) are required
to stabilize $X$ near the symmetric point.
\end{enumerate}
If these features are present,
this structure has predictive features:  gaugino masses are dominated by the anomaly mediated
contributions, while $B_\mu$ and $A$ terms are universal, $B_\mu = - m_{3/2} \mu$; $A=0$.
These models have possible
implications for the moduli problem.  Because the origin is a point of enhanced symmetry, it is natural that the
minimum of the $X$ potential lie at the origin, and that $X$  sit at the origin both immediately after inflation ends.
The latter point may be viewed as a virtue relative to the models of the previous section; alternatively, it is 
possible that anthropic issues related to light moduli select for the tuning needed in the supergravity models.
Finally, it should be noted, again, that in addition to the discrete $R$ symmetries, these models,
to be natural, require a discrete non-R symmetry.

\section{Conclusions:  Origins of Tuning}
\label{conclusions}

We have seen that supergravity models with metastable dynamical supersymmetry breaking are readily constructed
in the framework of retrofitting.  We have argued that discrete $R$ symmetries are likely to be an important feature
of these models, and we have focussed particularly on their consequences.  In the simplest models,
the goldstino supermultiplet, $X$, is neutral under the $R$ symmetry.  In these cases, we have seen that split supersymmetry
is not generic.   More generally, this framework is not particularly predictive; one can make
statements even about the $A$ and $B_\mu$ parameters only in restrictive circumstances.  We have seen that stable
supersymmetry breaking and the ISS models are similar in that, while gaugino mediation may dominate, additional elements
are required to understand $\mu$ and the cosmological constant.

We have considered the alternative possibility that the goldstino superfield is charged under the discrete $R$ symmetry or other symmetries.   The structure of the theory is distinctly more restricted,
and, for example, suppression of gaugino masses is automatic.  But understanding the smallness of the cosmological
constant requires unattractive features:  extremely small couplings or additional dynamics, introduced only
for this purpose.  Further fields and dynamics are necessary to
stabilize the moduli.  Somewhat more interesting is the possibility, which can be achieved in actual models, that $X$ is neutral
under the $R$ symmetry, but charged under some other discrete symmetry.  In this special set of circumstances,
split supersymmetry is automatic, there are predictions for $A$ and $B_\mu$, and
the superpotential is automatically of the correct order of magnitude to cancel the c.c.
Still, additional structure is required for moduli stabilization and there are generally
additional pseudo moduli.  Because of the additional structure required, this scenario does not appear generic.

The main issue with all of these theories is one of tuning.  Indeed,
it has long been argued that a high scale for supersymmetry breaking, of order $30$ TeV or so, would:
\begin{enumerate}
\item  Resolve the cosmological moduli problem
\item  Ameliorate the flavor problems of supersymmetry (including CP)
\end{enumerate}
A Higgs with mass of order $125$ GeV, it has been widely noted, would also point to such a scale.
But why a tuning of a part in $10^5$?  And if that large, why not larger.  
These questions might be related.  Recently, there have been suggestions that perhaps a large mass scale for the moduli
is an anthropic requirement.  The observed light element abundances have little anthropic significance, so if there is an anthropic
selection, it must arise for other reasons.  Possibilities include dark matter and formation of structure.  To address this, one needs a framework
capable of producing the observed baryon to photon ratio, dark matter density, baryon number density and perturbation spectrum, somewhere
within its parameter space.  Given such a model,
one can ask whether something like our observed universe is selected, with a high scale of moduli masses (supersymmetry
breaking).  This question is under study.

%\section{Conclusions}
%\label{conclusions}

%\bibliographystyle{hunsrt}
%\bibliographystyle{unsrt}
\bibliographystyle{jhep}
%\bibliographystyle{plain}
%\bibliographystyle{utphys}
%\bibliography{retrofitting_gravity_mediation}
%\bibliography{dinerefs}%{}
\bibliography{retrofitting_gravity_mediation.bbl}{}

\end{document}